# Dual-band bandpass filter derived from the transformation of a single-band bandpass filter

Chandramouli H. Mahadevaswamy
*Dept. of Electrical and Electronic*
*School of Engineering*
*University of Greenwich*
London, UK
ch9443g@greenwich.ac.uk

Anup Raju Vasistha
*Dept. of Electrical and Electronic*
*School of Engineering*
*University of Greenwich*
London, UK
ar9726r@greenwich.ac.uk

Augustine O. Nwajana
*Dept. of Electrical and Electronic*
*School of Engineering*
*University of Greenwich*
London, UK
a.o.nwajana@greenwich.ac.uk

*Abstract*—The recent proliferation of personal wireless communication devices is driving the need for multi-band frequency selective components including multiplexers and dual-band filters. This paper presents a simple technique for transforming a single-band bandpass filter (BPF) into a dual-band BPF. A second order (two-pole) single-band bandpass filter was chosen for this research, giving rise to a fourth order (four-pole) dual-band bandpass filter after the proposed filter transformation. Both filters were then implemented using the compact U-shaped microstrip resonator for improved device miniaturization. The proposed work features a centre frequency of 1.4 GHz for the single-band bandpass filter, with a span of 3.4% fractional bandwidth. The dual-band bandpass filter operates at 1.35 and 1.45 GHz. The design implementation employs the commercially available Rogers RT/Duroid 6010LM substrate, having a dissipation factor (tan δ) of 0.0023, dielectric constant (εr) of 10.7, diel thickness (h) of 1.27 mm, and top/bottom cladding of 35 microns. The results reported for the theoretical and practical designs show good agreement and improved performance when compared to similar research works in literature. The practical responses of the prototype dual-band BPF indicate a good return loss of better than 18 dB across both bands, and an insertion loss of better than 0.1 dB. The design prototype achieved physical size of 0.23 λg x 0.18 λg. The results reinforce the design's competitive edge in performance. λg is the guided wavelength for the microstrip line impedance at the centre frequency of the filter.

*Keywords—BPF; circuit conversion; coupling, microstrip; multi-band*

## I. Introduction

In the rapidly evolving landscape of communication technology, the demand for smaller and more efficient devices has become increasingly prevalent. This necessity has spurred research and development efforts towards enhancing the performance of primary devices, particularly filters, which play a critical role in signal transmission and reception. Filters are passive components designed to selectively transmit or receive signals within specified frequency bands [1]–[9]. They find applications across a myriad of communication systems. One remarkable advancement in this domain is the exploration of dual-band bandpass filter (BPF) as a single component in communication systems [10]–[18]. This cutting-edge approach aimed at achieving compactness and reduced losses in the system. Dual-band BPFs, realised by employing dual- and multi-mode resonators [19]–[22], have gained attention for their ability to significantly reduce the physical size of the filter device while enhancing overall performance. Resonators play a pivotal role in the investigation of BPF design techniques. Diverse resonator structures including hairpin [8], quarter-wavelength [1], open-loop ring [15], waveguide [20], substrate integrated waveguide [14], and folded-arms square open-loop [23], have been employed by researchers to successfully achieve BPF components.

In this paper, a simple technique for transforming a single-band bandpass filter into a dual-band bandpass filter is presented. The proposed method can be extended to any multi-band bandpass filter and implemented using any resonator of choice. Microstrip U-shaped resonator structure [24], [25] is employed in the implementation of the proposed prototype filters due to its simplicity and relative compact size, when compared to other numerous popular microstrip resonator structures. To realise the microstrip U-shaped resonator structure, the conventional microstrip half-wavelength resonator structure shown in Figure 1 (a) is folded into a U-shape structure with dimensions relative to the standard guided wavelength as shown in Figure 1 (b).

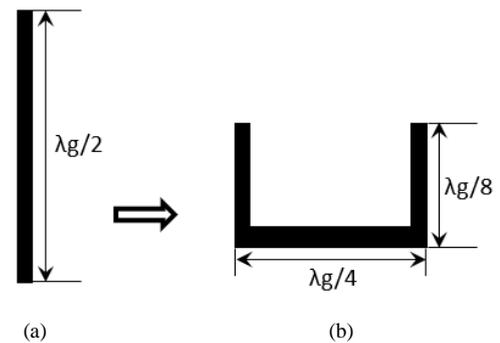

Fig. 1. Resonator configurations (a) half-wavelength, (b) U-shaped.

## II. Theoretical Filter Configuration

The procedure for achieving a single-band BPF begins with the transformation of the standard normalised Chebyshev lowpass prototype filter element-values of g0 = 1.0, g1 = 0.6648, g2 = 0.5445, and g3 = 1.2210 to realise a second order channel filter at the desired centre frequency of 1.4 GHz, fractional bandwidth of 3.4%, and the input/output characteristic impedance (Z0) of 50 Ohms. The single-band

BPF transformation is based on the principles reported in [25], and the final circuit configuration is shown in Figure 2(a). The single-band BPF is then converted into the corresponding dual-band BPF configuration shown in Figure 2(b). The transformation from the single-band to the dual-band BPF is achieved by simply extending the circuit and retaining a similar mutual coupling inductor-only network between each pair of resonators as shown in Figure 2(b). The numerical values for the capacitive and the inductive parameters indicated in Figure 2 are C = 44.4564 pF, L = 0.2907 nH, L01 = 5.6841 nH, L12 = 5.1442 nH. The numerical values are all based on the filter design formulations previously reported in [23].

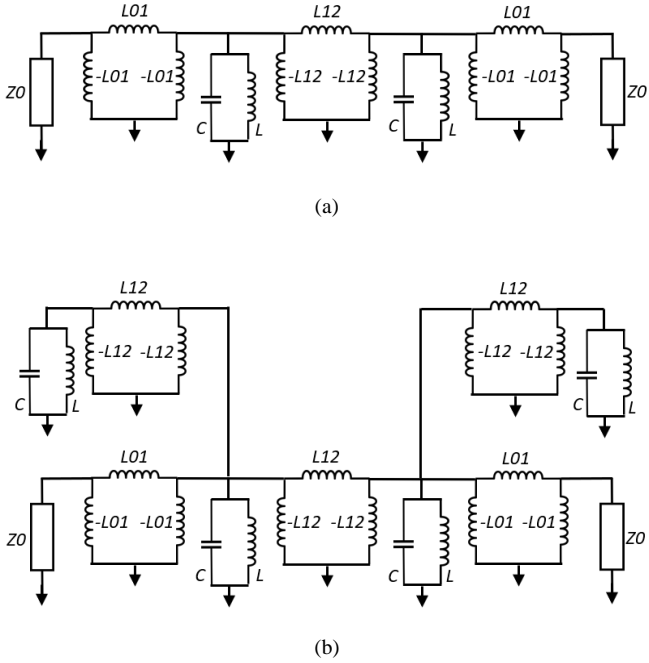

Fig. 2. Filter circuit configurations (a) single-band, (b) dual-band.

## III. PRACTICAL FILTER CONFIGURATION

The microstrip U-shaped resonator structure reported in [24], [25] is employed in the implementation of the proposed prototype filters due to its simplicity and relative compact size, when compared to other numerous popular microstrip resonator structures. The width and guided wavelength for designing the U-shaped resonator are determined from [23]. PathWave Advanced Design System (ADS) EM Simulation Software is used to construct and simulate the layout of both filter models, with the coupling arrangements shown in Figure 3, forming the design foundation. The layouts were created on Rogers RT/Duroid 6010LM substrate with a dielectric constant of 10.7, a thickness of 1.27 mm, and a loss tangent of 0.0023. The microstrip U-shaped resonator is designed to resonate at the filter circuits specified centre frequency of 1.4 GHz.

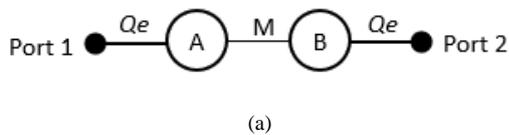

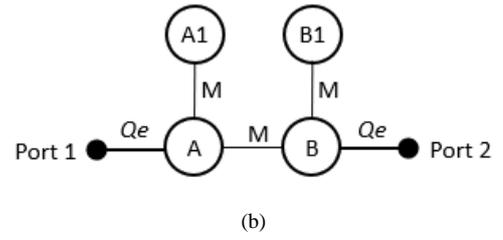

Fig. 3. Filter coupling arrangements (a) single-band, (b) dual-band.

$$M = \frac{FBW}{\sqrt{g_1 g_2}} = 0.0565 \quad (1)$$

$$Q_e = \frac{g_0 g_1}{FBW} = 19.5529 \quad (2)$$

Equations (1) and (2) proposed in [24] were employed in the determination of the mutual coupling coefficient, M, between two adjacent resonators (A and B, A and A1, B and B1); and the external quality factor, $Q_e$, between the input port and the first filter resonator (that is, port 1 and A), and the output port and the last filter resonator (that is, port 2 and B), respectively. The physical dimensions of the filters microstrip layouts, and the corresponding images of the fabricated filter devices are shown in Figures 4 and 5, respectively.

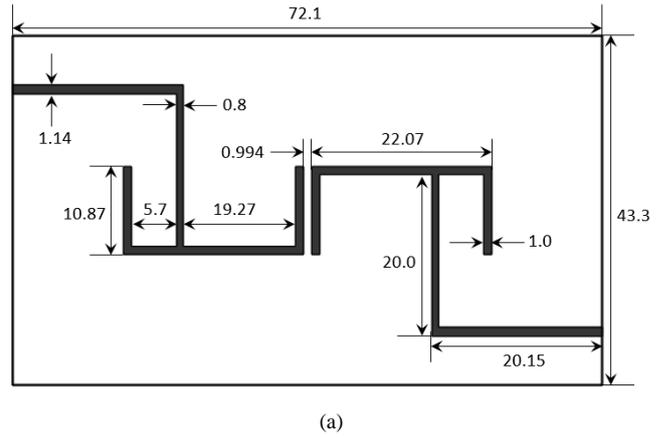

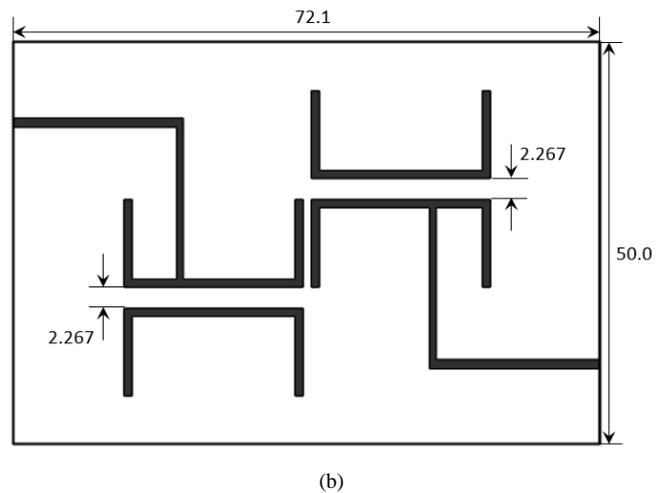

Fig. 4. Microstrip filter layouts with dimensions in mm (a) single-band, (b) dual-band.

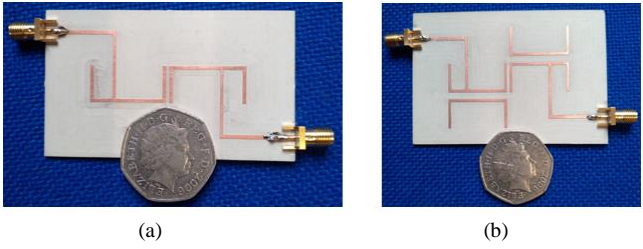

(a)             (b)

Fig. 5. Fabricated filter image (a) single-band, (b) dual-band.

## IV. RESULTS AND DISCUSSION

This section captures and discusses both the theoretical and practical results for the proposed single- and dual-band bandpass filters. The dual-band bandpass filter implemented using the microstrip U-shaped resonator has been developed from the transformation of a single-band bandpass filter. The results for the single- and dual-band bandpass filters are presented in Figure 6 for ease of analysis and comparison. The results show good agreements between the corresponding theoretical and practical responses. Looking at Figure 6(a), the single-band bandpass filter transmits at the specified centre frequency of 1.4 GHz, with an insertion and return losses of 0.1 dB and 20 dB, respectively. In Figure 6(b), the dual-band bandpass filter transmits at 1.35 GHz and 1.45 GHz, with insertion losses of better than 0.1 dB across both bands. The return loss achieved in the first band is better than 24 dB, while that of the second band is better than 18 dB. The proposed dual-band bandpass filter results are compared to the current state-of-the-art and captured in Table 1. The performance comparison reinforces the design's competitive edge over similar designs reported in literature. The designs achieved a physical size of 0.23 $\lambda_g$ x 0.15 $\lambda_g$ for the single-band bandpass filter, and 0.23 $\lambda_g$ x 0.18 $\lambda_g$ for the dual-band bandpass filter. It is important to note that the reported results assume copper conductor with cladding thickness of 35 micron (µm) and conductivity of 5.8 x $10^7$ S/m. Thickness variation of the substrate material used in the design, and metal surface roughness were not considered in this research work.

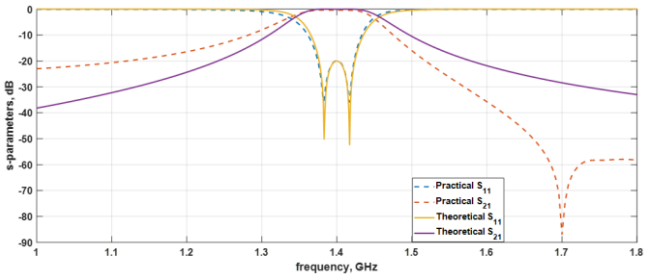

(a)

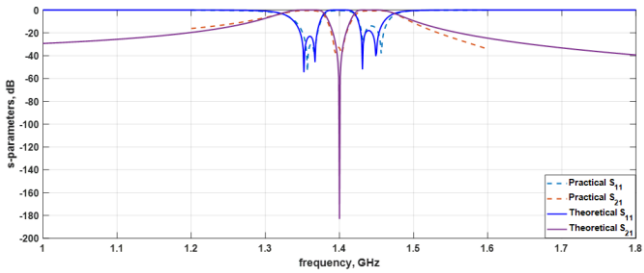

(b)

Fig. 6. Theoretical and practical filter results comparison (a) single-band, (b) dual-band.

TABLE I. PERFORMANCE COMPARISON WITH RELATED LITERATURE

| Ref. | $f_1/f_2$ (GHz) | Filter order | Size $(\lambda_g)^2$ | TL[a] | RL[b] (dB) | IL[c] (dB) |
|---|---|---|---|---|---|---|
| [1] | 2.3/3.7 | 2/2 | 0.0224 | MS | 23/21 | 1.1/1.9 |
| [7] | 7.8/12.3 | 2/2 | 0.0156 | Hybrid | 15/16 | 1.3/1.5 |
| [10] | 0.9/2.5 | 2/2 | 0.0130 | MS | 20/12 | 0.3/0.9 |
| [12] | 2.9/4.4 | 2/2 | 0.0308 | MS | 17/17 | 0.1/0.1 |
| [15] | 1.6/4.7 | 2/2 | 0.0410 | SIW | 14/21 | 1.1/1.2 |
| [18] | 1.1/3.4 | 2/3 | 0.0221 | Hybrid | 14/14 | 0.4/1.3 |
| [20] | 2.9/3.5 | 2/2 | 0.0306 | WG | 17/18 | 1.0/1.5 |
| [21] | 42.8/60.5 | 4/6 | 1.8696 | SIW | 18/12 | 1.9/1.4 |
| This work | 1.35/1.45 | 2/2 | 0.0410 | MS | 24/18 | 0.1/0.1 |

[a] transmission line, [b] return loss, [c] insertion loss.

## V. CONCLUSION

A dual-band bandpass filter achieved from the transformation of a single-band bandpass filter has been developed and realized using microstrip U-shaped resonator technology. The proposed dual-band bandpass filter operates at 1.35 and 1.45 GHz centre frequencies. The realization and fabrication are carried out on Rogers RT/Duroid 6010LM substrate with a dielectric constant of 10.7, a thickness of 1.27 mm and a loss tangent of 0.0023. The theoretical and practical results of the proposed dual-band bandpass filter show good agreement with a return loss of better than 18 dB across the two bands, with an insertion loss of better than 0.1 dB on each band. The fabricated dual-band bandpass filter device achieved a physical size of 0.23 $\lambda_g$ x 0.18 $\lambda_g$. Looking at the performance comparison in Table 1, it is easy to conclude that the proposed dual-band bandpass filter competes well with the current state-of-the-art as it has the best insertion loss and a very good return loss as recorded in Table 1. The proposed dual-band bandpass filter device finds application in modern multifunctional communication systems where it is used for isolating small sections of a frequency spectrum within a wider spectrum.

## REFERENCES

[1] M. Ma, F. You, T. Qian, C. Shen, R. Qin, T. Wu, and S. He, "A wide stopband dual-band bandpass filter based on asymmetrical parallel-coupled transmission line resonator," *IEEE Transactions on Microwave Theory and Techniques*, vol. 70, no. 6, pp. 3213–3223, Jun. 2022.

[2] Y. C. Li, D.-S. Wu, T.-Z. Zhang, and Q. Xue, "Dual-band dual-channel bandpass filter using high quality factor dielectric resonator," *IEEE Transactions on Circuits and Systems II: Express Briefs*, vol. 70, no. 6, pp. 1931–1935, Jun. 2023.

[3] A. O. Nwajana, E. R. Obi, G. K. Ijemaru, E. U. Oleka, and D. C. Anthony, "Fundamentals of RF/microwave bandpass filter design," Handbook of Research on 5G Networks and Advancements in Computing, Electronics, and Electrical Engineering, IGI Global: Hershey, PA, USA, 2021.

[4] A. O. Nwajana, "A step-by-step approach to bandpass/channel filter design," *International Journal of Electronics, Communications, and Measurement Engineering*, vol. 10, no. 2, pp. 1–14, Jul. 2021.

[5] Z. Yan, X. Zhang, X. He, Y. Ye, B. Hou, C. Hu, and W. Wen, "A high-performance dual-band switchable bandpass filter and its application in 5G signal detector," *IEEE Transactions on Components, Packaging and Manufacturing Technology*, vol. 13, no. 8, pp. 1242–1253, Aug. 2023.

[6] R. Mu, Y. Wu, L. Pan, W. Zhao, and W. Wang, "A miniaturized low-loss switchable single- and dual-band bandpass filter," *International*

[7] L. Gu, and Y. Dong, "Compact dual-band bandpass filter with high selectivity based on spiral-stub-loaded dual-mode QMSIW resonators," *Microwave and Optical Technology Letters*, vol. 65, pp. 559–566, Oct. 2023.

[8] A. O. Nwajana, "Circuit modelling of bandpass/channel filter with microstrip implementation," *Indonesian Journal of Electrical Engineering and Informatics*, vol. 8, no. 4, pp. 696–705, Dec. 2020.

[9] A. O. Nwajana, A. Dainkeh, and K. S. K. Yeo, "Substrate integrated waveguide (SIW) bandpass filter with novel microstrip-CPW-SIW input coupling," *Journal of Microwaves, Optoelectronics and Electromagnetic Applications*, vol. 16, no. 2, pp. 393–402, Jun. 2017.

[10] G. Y. Wei, Y. X. Wang, J. Liu, and H. P. Li, "Design of a planar compact dual-band bandpass filter with multiple transmission zeros using a stub-loaded structure," *Progress in Electromagnetic Research Letters*, vol. 109, pp. 23–30, Feb. 2023.

[11] J.-M. Yan, B.-J. Kang, Y. Yang, and L. Cao, "Single- and dual-band bandpass filters based on a novel microstrip loop-type resonator loaded with shorted stubs," *Progress in Electromagnetic Research Letters*, vol. 113, pp. 61–67, Nov. 2023.

[12] T.-H. Lee, K.-C. Yoon, and K. G. Kim, "Miniaturized dual-band bandpass filter using T-shaped line based on stepped impedance resonator with meander line and folded structure," *Electronics*, vol. 11, no. 219, pp. 1–8, Jan. 2022.

[13] H. Liu, J. Kuang, Y. Cao, and Y. Zhang, "Flexible design of dual-band common-mode filters using hairpin ring resonators and defected ground slots," *IEEE Transactions on Electromagnetic Compatibility*, vol. 65, no. 5, pp. 1360–1370, Oct. 2023.

[14] A. O. Nwajana, and E. R. Obi, "A review on SIW and its applications to microwave components," *Electronics*, vol. 11, no. 1160, pp. 1–21, Apr. 2022.

[15] N. C. Pradhan, S. Koziel, R. K. Barik, A. Pietrenko-Dabrowska, and S. S. Karthikeyan, "Miniaturized dual-band SIW-based bandpass filters using open-loop ring resonators," *Electronics*, vol. 12, no. 3974, pp. 1–15, Sep. 2023.

[16] A. J. Alazemi, "Dual-band and wideband bandpass filters using coupled lines and tri-stepped impedance stubs," *Micromachines*, vol. 14, no. 1254, pp. 1–14, Jun. 2023.

[17] K. V. P. Kumar, V. K. Velidi, A. A. Althuwayb, and T. R. Rao, "Microstrip dual-band bandpass filter with wide bandwidth using paper substrate," *IEEE Microwave and Wireless Components Letters*, vol. 31, no. 7, pp. 833–836, Jul. 2021.

[18] X.-F. Li, and J.-K. Xiao, "Dual-band bandpass filter based on suspended coplanar waveguide-microstrip hybrid," *IEEE Transactions on Circuits and Systems I: Regular Papers,*" vol. 70, no. 10, pp. 3920–3929, Oct. 2023.

[19] K. S. K. Yeo, and A. O. Nwajana, "A novel microstrip dual-band bandpass filter using dual-mode square patch resonator," *Progress in Electromagnetic Research C*, vol. 36, pp. 233–247, Jan. 2013.

[20] Z. Xu, Y. Wu, Q. Dong, and W2. Wang, "Miniaturized dual-band filter using dual-mode dielectric waveguide resonator," *IEEE Microwave and Wireless Components Letters*, vol. 32, no. 12, pp. 1411–1414, Dec. 2022.

[21] X.-L. Yang, X.-W. Zhu, and X. Wang, "Dual-band substrate integrated waveguide filters based on multi-mode resonator overlapping mode control," *IEEE Transactions on Circuits and Systems II: Express Briefs*, vol. 70, no. 6, pp. 1971–1975, Jun. 2023.

[22] Q. Liu, D.-W. Zhang, K. Gong, H. Qian, W.-Z. Qu, and N. An, "Single- and dual-band bandpass filters based on multiple-mode folded substrate-integrated waveguide cavities," *IEEE Transactions on Microwave Theory and Techniques*, vol. 71, no. 12, pp. 5335–5345, Dec. 2023.

[23] A. O. Nwajana, and E. R. Obi, "Application of compact folded-arms square open-loop resonator to bandpass filter design," *Micromachines*, vol. 14, no. 320, pp. 1–10, Jan. 2023.

[24] [24] J.-S. Hong, *Microstrip Filters for RF/Microwave Applications*, 2nd ed.; John Wiley & Sons: Hoboken, NJ, USA, 2011.

[25] M. A. Saeed, and A. O. Nwajana, "U-shaped terahertz microstrip patch antenna for 6G future communications," in Proceedings of the 2023 7th International Electromagnetic Compatibility Conference (EMC Turkiye), Istanbul, Turkiye, 17–20 September 2023, pp. 1–4.